# Expanding the Materials Search Space for Multivalent Cathodes

Ann Rutt,[a] Jimmy-Xuan Shen,[a] Matthew Horton,[b] Jiyoon Kim,[a] Jerry Lin,[a] and Kristin A. Persson [ab]

a. Department of Materials Science and Engineering, University of California, Berkeley, USA.

b. Materials Sciences Division, Lawrence Berkeley National Laboratory, Berkeley, USA.

## Abstract

Multivalent batteries are an energy storage technology with the potential to surpass lithium-ion batteries, however their performance has been limited by the low voltages and poor solid-state ionic mobility of available cathodes. A computational screening approach to identify high-performance multivalent intercalation cathodes among materials that do not contain the working ion of interest has been developed which greatly expands the search space that can be considered for materials discovery. This approach has been applied to magnesium cathodes as a proof of concept and four resulting candidate materials (NASICON $V_2(PO_4)_3$, birnessite $NaMn_4O_8$, tavorite $MnPO_4F$, and spinel $MnO_2$) are discussed in further detail. In examining the ion migration environment and associated $Mg^{2+}$ migration energy in these materials, local energy maxima are found to correspond with pathway positions where $Mg^{2+}$ passes through a plane of anion atoms. While previous works have established the influence of local coordination on multivalent ion mobility, these results suggest that considering both the type of local bonding environment as well as available free volume for the mobile ion along its migration pathway can be significant for improving solid-state mobility.

## Introduction

Multivalent (e.g. Mg, Ca, Zn) batteries have been explored as a "beyond Li-ion" technology but are currently limited in their promise as high-energy-density batteries by the lack of suitable electrolytes as well as cathodes.[1–3] Among the most promising recent advancements in Mg cathodes are spinel $Mg_xTi_2S_4$[4,5] and layered $Mg_xTiS_2$[4,6] which show improved capacities (160 mAh/g for spinel $Mg_xTi_2S_4$, 140 mAh/g for layered $Mg_xTiS_2$) compared to the original Mg cathode, Chevrel $Mg_xMo_6S_8$ (100 mAh/g).[7] However, the spinel $Mg_xTi_2S_4$ and the layered $Mg_xTiS_2$ cathodes share two limitations that must be overcome in order to realize high-performance magnesium batteries: 1) low voltages (1.2 V vs. $Mg/Mg^{2+}$) and 2) poor solid-state mobility (requiring elevated cycling temperatures of 60°C).

Calcium battery research is still in its early stages especially in comparison to Mg-based systems, but there have been recent advances in reversible Ca plating.[8–10] Some promising Ca cathode candidates such as $Na_{0.5}VPO_{4.8}F_{0.7}$ (NVPF)[11] and the sodium superionic conductor (NASICON) $NaV_2(PO_4)_3$[12,13] have shown capacities of 87 mAh/g and 83 mAh/g respectively. Compared to their Mg counterparts, these compounds show higher voltages of ~ 3.2 V vs. $Ca/Ca^{2+}$ and slighter lower $Ca^{2+}$ intercalation capacities. Elevated operation temperatures of 75°C also tend to improve capacities. It has proved difficult to identify a high voltage and high capacity Ca cathode that performs well in ambient conditions with repeated cycling.

The challenge of identifying multivalent intercalation cathodes with good solid-state mobility correlates directly to their potential for higher capacity.[1] While the higher valence of multivalent ions can lead to a higher energy density, there is also a trade-off associated with poor mobility due to the stronger Coulombic interactions between the mobile multivalent ion and surrounding cathode host lattice. Previous work[14] has shown that multivalent ion mobility correlates with the local coordination topology along the diffusion path, favoring flat electrostatic landscapes with no strong binding sites for the multivalent ion. For example, materials which contain Mg in their as-synthesized states tend to exhibit relatively strong Mg binding sites, and hence poorer mobility.

Interestingly, the most successful Mg cathodes have originated from compounds synthesized in a form without Mg. In the case of Chevrel $Mg_xMo_6S_8$, the copper-containing $CuMo_3S_4$ is synthesized and the copper then chemically removed.[15] Similarly, for spinel $Mg_xTi_2S_4$, $CuTi_2S_4$ is synthesized and then the copper removed by oxidation.[5] Layered $Mg_xTiS_2$ can be synthesized free of any intermediate compound as $TiS_2$.[6] The Ca cathodes discussed previously were also originally synthesized without Ca. $Na_{1.5}VPO_{4.8}F_{0.7}$ and NASICON $Na_3V_2(PO_4)_3$ are synthesized and then partially electrochemically desodiated to obtain $Na_{0.5}VPO_{4.8}F_{0.7}$ and $NaV_2(PO_4)_3$,[11–13] which can cycle as Ca cathodes. A successful multivalent cathode which has been synthesized in the discharged state (where the multivalent ion is contained in the structure) has yet to be reported.[1]

Hence, one may extrapolate that identifying new multivalent cathodes among materials that already contain the multivalent ion of interest would yield scarce results. An automated computational infrastructure for discovering intercalation electrodes has been previously reported,[16] however these efforts have focused exclusively on evaluating materials that already contain Mg and where the intercalation sites are already known. Given the solid-state mobility challenges with multivalent ions, new strategies are needed to discover high-performance multivalent cathodes. In this paper we report a new comprehensive computational framework for identifying novel Mg cathode materials in compounds that do not a priori contain Mg, e.g. are most stable in their charged state. We demonstrate our approach to computationally screen thousands of materials from the Materials Project database[17] in order to evaluate promising Mg cathodes.

## Methods

The presented computational approach for evaluating materials free of magnesium as Mg cathodes can be described in four screening tiers. The criteria for each tier were ordered by a combination of robustness and computational cost. Less computationally demanding methods, which still correlate well to experimental results, were used first in earlier tiers in order to limit applying more expensive methods to a smaller number of candidate materials. Each tier focused on a different set of material properties (and are described in more detail):
1. Relative stability and composition
2. Reducible specie oxidation state
3. Insertion site finding
4. Multivalent ion solid-state mobility

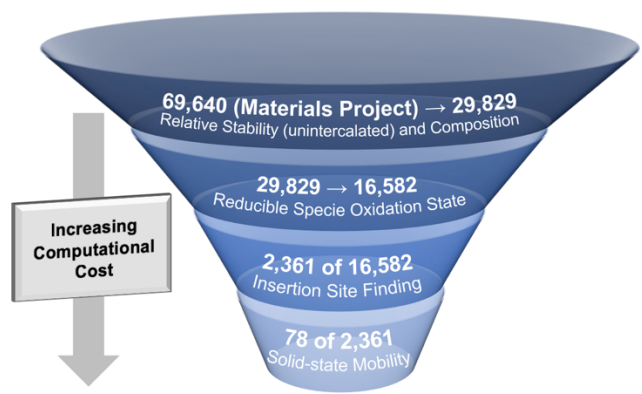

Figure 1. A funnel graphic summarizing the screening process for finding promising Mg cathodes in materials free of magnesium. The screening process has been divided into four stages in order of increasing computational cost.

**Relative Stability and Composition.** The Materials Project, a database of density functional theory (DFT) calculations of inorganic crystals and molecules, was used to source candidate materials.[17] We note that the examination of Mg-free ("empty") hosts presents a vast space of many tens of thousands of possible materials. Two properties, relative stability and composition, were used to narrow down this search space. Relative stability compared to other materials composed of the same elements was captured by the quantity, energy per atom above the convex hull, which encloses the most stable phases within the relevant chemical space.[18,19] This "energy above hull" is 0 eV/atom for a material that is predicted to be thermodynamically stable at 0 K. A cut-off value of < 0.2 eV/atom was used to select compounds that are likely to be synthesizable based on previous work that established sensible cut-off values.[20,21] Composition constraints were also applied that required the candidate material: 1) to contain at least one redox active element (Ti, V, Cr, Mn, Fe, Co, Ni, Cu, Nb, Mo, Ru, Ag, W, Re, Sb, Bi), 2) to contain either oxygen or sulfur, and 3) excludes radioactive elements (elements with an atomic number between 84 and 104). Using these criteria, the Materials Project database (version 2018.11), which contained 69,640 inorganic crystals at the time that this work was started, was reduced to 29,829 candidate materials of interest.

**Reducible Specie Oxidation State.** Given the intention to insert magnesium into empty candidate host materials, viable candidates need to encompass a reducible specie that can accept electrons upon Mg insertion. The next screening tier is focused on finding materials with a high enough oxidation state to permit reduction. There are two algorithms within pymatgen[22] that are able to suggest likely oxidation states of a given material. One uses a bond valence method[23] to determine oxidation states and the other predicts appropriate oxidation states based on the material's chemical formula using a data mining approach. Candidate materials were discarded at this tier if they did not contain any of the following reducible species: $Ti^{4+}$, $V^{4+,5+}$, $Cr^{4+,5+,6+}$, $Mn^{3+,4+,5+,6+,7+}$, $Fe^{3+,4+,5+,6+}$, $Co^{3+,4+,5+}$, $Ni^{3+,4+}$, $Cu^{2+,3+,4+}$, $Nb^{5+}$, $Mo^{4+,5+,6+}$, $Ru^{5+,6+,7+,8+}$, $Ag^{2+,3+}$, $W^{6+}$, $Re^{7+}$, $Sb^{5+}$, $Bi^{4+,5+}$. Both methods were applied to evaluate oxidation states and in the case of disagreeing results, the bond valence method was preferred. Screening by reducible specie oxidation state narrowed the candidates of interest from 29,829 to 16,582 materials.

**Insertion Site Finding.** The next tier of screening candidate materials considers the identification of Mg insertion sites. Recently, Shen et al. reported a new approach for identifying the location of insertion sites in any given crystal structure which will be referred to as the "insertion algorithm".[24] The insertion algorithm uses the calculated charge density of the material to identify charge density minima which were shown to correlate strongly with viable insertion sites in known electrode materials. For each possible insertion site, the working ion of interest (such as Mg in this case) is inserted with one working ion per unit cell. A DFT relaxation is then performed to refine the site location and evaluate the change in lattice and crystal structure to assess the possibility of topotactic insertion (i.e. whether the host structure is retained after insertion). As this tier requires multiple DFT calculations per material, it was not viable to apply the insertion algorithm to all 16,582 candidate materials at the time this work was performed. To demonstrate the screening workflow, the insertion algorithm was applied to a random selection of 2,361 materials. Given the investigation of empty cathodes that are more stable in their charged state, the insertion algorithm was set to explore the possibility of a single working ion per unit cell, as a first, necessary requirement. Further successive working ion insertions can be repeated until the insertion is no longer topotactic or the minimum redox state of the compound is reached, thus determining the maximum intercalation level.

Of the 2,361 materials where the insertion algorithm was applied, 1,767 were discarded because none of the single magnesium insertions attempted were topotactic (i.e. the host structure changed significantly upon insertion). Additional properties useful for screening become computable after completing the insertion algorithm for a material. These properties include the voltage and relative stability (energy above hull) of the partially/fully magnesiated compound. The remaining 594 candidate materials were prioritized for the last tier of evaluating $Mg^{2+}$ solid-state mobility using the following criteria: 1) average voltage > 1.5 V, 2) energy per atom above the convex hull of the charged (unmagnesiated) material < 0.05 eV/atom, and 3) energy per atom above the convex hull of the discharged (magnesiated) material < 0.1 eV/atom. After applying these criteria, the number of candidates was further reduced by using the structure matching capabilities in pymatgen[22] to select one material that would be representative of each unique structure type. As a result, 78 materials were selected for the next tier.

Applying the insertion algorithm to a host material produces a list of valid topotatic insertion sites for Mg. These sites can be used to form a migration graph representing an interconnected network of Mg sites in the material as introduced in our previous work.[25] Sets of neighboring Mg sites can be extracted from this migration graph which correspond to a segment of a possible $Mg^{2+}$ migration pathway in the material. Images representing the $Mg^{2+}$ in various positions along these pathway segments can be generated and paired with DFT calculations in order to evaluate the energetics along the pathway segment. This information provides a key input for evaluating $Mg^{2+}$ solid-state mobility in a given material, which is addressed in the following section.

**Multivalent Ion Solid-State Mobility.** The last screening tier estimates the minimum energy barrier required for $Mg^{2+}$ to migrate through the material and corresponds to the most computationally expensive tier. An upper limit migration barrier of 650meV would remove materials which exhibit sluggish intrinsic ionic mobility and provide a possibility for good rate capability that might allow for a C/2 cycling rate with nanosized particles.[14] Given the evaluation of Mg-free compounds, only mobility in the charged (deintercalated) state at the dilute lattice limit was considered. Supercells were generated using pymatgen[22] to avoid fictious self-interaction effects from a neighboring $Mg^{2+}$ due to periodic boundary conditions. Finally, the ApproxNEB algorithm[26] implemented through the python package, atomate,[27] was used to evaluate the migration barrier for a given pathway segment. The ApproxNEB algorithm was selected over the traditional Nudged Elastic Band (NEB) scheme due to its lower computational cost and robustness which makes it more appropriate for high-throughput applications.[26] Implemented in atomate, the ApproxNEB algorithm performs a series of relaxations for host, end point, and image structures for the specified migration events in a material. The key difference between NEB and ApproxNEB is in how the image relaxations are handled. With ApproxNEB, the images are relaxed independently of each other, and a coherent mobile ion path is maintained by fixing the positions of 2 atoms (the mobile working ion and the atom furthest away) in each image relaxation. Given the constraints of the ApproxNEB image relaxations, this method is likely to provide a slight overestimation of the energy barrier as compared to NEB. The energies produced by the ApproxNEB algorithm were mapped back onto the connections in the migration graph for a material. Pathway segments with incomplete points where calculations failed to reach sufficient convergence were excluded. The migration graph populated with the available energy information was used to locate the migration pathway through a material with the lowest barrier.

**ApproxNEB Calculation Details.** The Vienna Ab initio Software Package (VASP) was used to perform density functional theory (DFT) calculations where the exchange correlation was approximated with the Perdew–Burke–Ernzerhof (PBE) generalized gradient approximation (GGA). Pseudopotentials were selected according to "MPRelaxSet" specified in pymatgen.[22] A U term was not included in these calculations as there is no conclusive evidence that GGA+U performs better when investigating ion migration with methods such as NEB.[28–31] The total energy was sampled using a Monkhorst-Pack mesh with k-point density of 64 Å$^{-3}$. Projector augmented-wave theory combined with a well-converged plane-wave cutoff of 520 eV were used to describe the wave functions. The convergence threshold of the total energy was set to 0.0005 eV and a force tolerance of 0.05 eV/Å.

## Results

From the data set generated by the described screening process, 14 candidate materials were found to exhibit viable pathways for $Mg^{2+}$ migration (ApproxNEB estimated barriers < 800meV). A few of these materials have been identified as novel and are currently being pursued electrochemically. Four materials are highlighted here as candidate Mg cathodes to illustrate the value of this screening approach: $V_2(PO_4)_3$ (mp-26962), $NaMn_4O_8$ (mp-1016155), $MnPO_4F$ (mp-25426), and $MnO_2$ (mp-25275) which are depicted in Figure 2. Table 1 summarizes the following properties of these cathodes: voltage, capacity (upon a single Mg insertion), charge (deintercalated) stability, discharge (intercalated) stability, and the estimated ApproxNEB energetic barrier for $Mg^{2+}$ migration. Figure 3 shows the migration energy landscape determined with the ApproxNEB algorithm for $Mg^{2+}$ along the best percolating path identified.

| Material | Voltage | Capacity | Charge (Deintercalated) Stability | Discharge (Intercalated) Stability | Energetic Barrier (dilute lattice limit) |
|---|---|---|---|---|---|
| Monoclinic NASICON $Mg_{(x<0.5)}V_2(PO_4)_3$ mp-26962 | 3.3 V | 67 mAh/g | 23 meV/atom | -6 meV/atom | 671 meV |
| Sodiated Birnessite $Mg_{(x<1)}NaMn_4O_8$ mp-1016155 | 2.2 V | 136 mAh/g | 19 meV/atom | 82 meV/atom | 200 meV |
| Tavorite $Mg_{(x<0.5)}MnPO_4F$ mp-25426 | 3.7 V | 148 mAh/g | 42 meV/atom | 62 meV/atom | 1015 meV |
| Spinel $Mg_{(x<0.5)}MnO_2$ mp-25275 | 2.9 V | 144 mAh/g | 51 meV/atom | 45 meV/atom | 711 meV |

Table 1. A summary of electrode properties for the four Mg cathodes that will be described in more detail. The voltage reported is a theoretical voltage (V) calculated from the energy difference of the intercalation reaction ($\Delta G_{rxn} = -nFV$). The theoretical capacity (Q) was calculated based on the atomic mass of the intercalated material (M) with $Q = nF/M$. n represents the number of electrons transferred (for Mg, n=2) and F is Faraday's constant. Stability values (energy per atom above the convex hull) were calculated using the MP2020Compatibility scheme[32] and Materials Project database phase diagrams using pymatgen.[22] The lowest energetic barrier for $Mg^{2+}$ migrating along a percolating pathway calculated with ApproxNEB at the dilute lattice limit (single Mg in host material supercell) is listed.

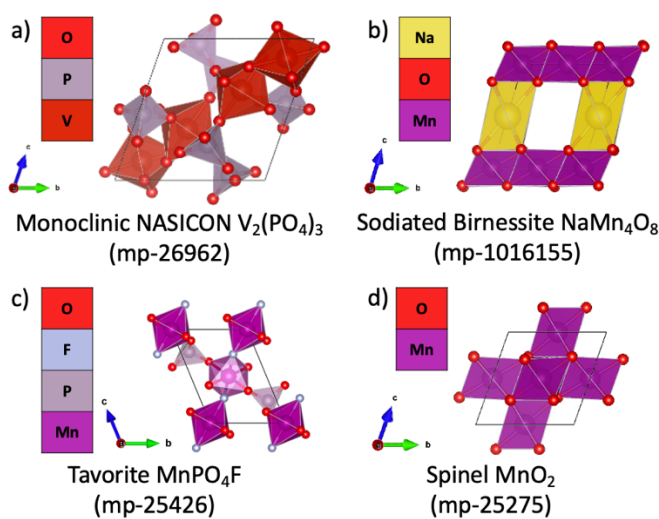

Figure 3. Unit cell crystal structures of each highlighted material in their unintercalated form.

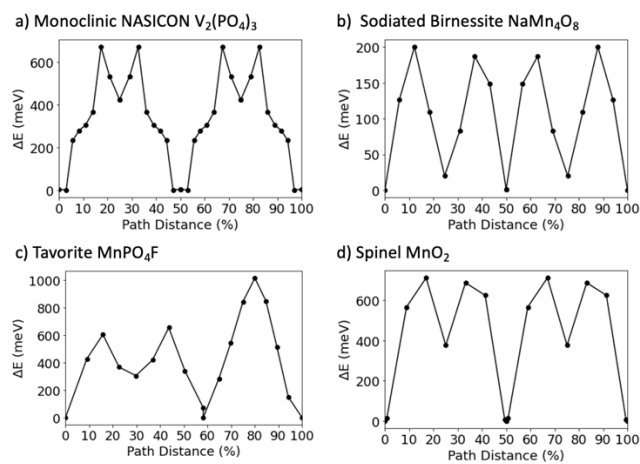

Figure 2. Plots of the energy landscape of $Mg^{2+}$ migration along a percolating path for the 4 candidate materials. The percolating path starts from a given Mg site in a unit cell to the equivalent site in a neighboring unit cell and the energy evolution along the path is predicted with ApproxNEB at the dilute lattice limit (single Mg in host material supercell). Pathways were reduced to symmetrically distinct hops which were calculated separately and then mapped back to the pathway for computational efficiency by avoiding redundant calculations. Lines connecting adjacent points are provided as a guide to the eye. a) Monoclinic NASICON $V_2(PO_4)_3$ (mp-26962) has an energetic barrier of 671 meV along a total path distance of 13.2 A. b) Sodiated Birnessite $NaMn_4O_8$ (mp-1016155) has an energetic barrier of 200 meV along a total path distance of 6.8 A. c) Tavorite $MnPO_4F$ (mp-25426) has an energetic barrier of 1015 meV along a total path distance of 7.8 A. d) Spinel $MnO_2$ (mp-25275) has an energetic barrier of 711 meV along a total path distance of 7.4 A.

**$V_2(PO_4)_3$ (mp-26962)** is a monoclinic NASICON that can be obtained experimentally from its lithiated version, $Li_3V_2(PO_4)_3$. This material has been studied as a Mg cathode where XAS spectra showed a change in vanadium oxidation state upon Mg intercalation,[33] however definitive evidence of reversible Mg intercalation with repeated cycling has yet to be reported.[34] The voltage predicted by the screening process (3.3 V) compares well with the experimentally measured value of ~3.0 V vs. $Mg/Mg^{2+}$. Due to the single Mg insertion explored here, the calculated capacity corresponds to a lower limit magnesiation level of $Mg_{(x<0.5)}V_2(PO_4)_3$ and hence is below the experimentally reported capacity of ~197 mAh/g for $Mg_{(x<1.5)}V_2(PO_4)_3$. Sufficient $Mg^{2+}$ mobility for acceptable rate capability appears possible with an ApproxNEB predicted migration barrier of 671 meV. While this value is greater than the 650 meV threshold, as stated previously, ApproxNEB is known for overestimating the energy barrier; hence it is likely the true activation energy for dilute $Mg^{2+}$ migration is lower. The results of this screening recommends further investigation of $V_2(PO_4)_3$ as a Mg cathode, particularly highlighting the need for testing in electrolytes that are stable at sufficiently high voltages.

**$NaMn_4O_8$ (mp-1016155)** is a sodiated version of birnessite, $\delta$-$MnO_2$, which is the layered polymorph of $MnO_2$. Birnessite is typically hydrated with water molecules between layers of $MnO_6$ octahedra and has been studied in the literature as a Mg cathode.[34–36] More success has been found using aqueous electrolytes where there is a reversible transformation from $\delta$-$MnO_2$ to $\lambda$-$MnO_2$ upon discharge.[35] Involving water presents a challenge for using birnessite in high-energy-density Mg batteries given the incompatibility of water with magnesium metal anodes. However, pillaring the structure with sodium ions instead of water molecules avoids these compatibility issues and may help explain the good $Mg^{2+}$ mobility predicted by a low ApproxNEB migration barrier of 200 meV. Alkali-ion pillaring has been examined as a strategy to improve electrochemical performance in vanadium oxides as Li-ion cathodes by stabilizing the structure and improving ion mobility.[37–39] More detailed studies would be required to understand the role of sodium and possible avenues for facilitating multivalent ion mobility such as in NASICONs where sodium reordering upon calcium transport has been demonstrated.[12,13,40] While synthesis of $NaMn_4O_8$ has not been reported, replacing water with sodium in the birnessite structure could offer an avenue for using the layered $\delta$-$MnO_2$ polymorph as a cathode with magnesium metal anodes.

**MnPO$_4$F (mp-25426)** belongs to the tavorite structure family. This class of materials has been studied as Li-ion cathodes and includes examples such as LiVPO$_4$F,[41] LiFePO$_4$F,[42] and LiFeSO$_4$F.[43] Theoretical work on FeSO$_4$F[44] and VPO$_4$F[45] as Mg cathodes has been reported and found promising properties.[34] However, no Mg$^{2+}$ electrochemical experimental work has yet been published on these materials. With voltages of 2.5 V for FeSO$_4$F and 2.6 V for VPO$_4$F, perhaps difficulties in identifying higher voltage magnesium electrolytes present a roadblock. With a predicted voltage of 3.7 V for MnPO$_4$F, the lack of suitable electrolytes is also a limitation that must be overcome before this material can be pursued experimentally. While reasonable Mg$^{2+}$ migration barriers have been identified for other tavorites (360 meV for FeSO$_4$F and 704 meV for VPO$_4$F), the dilute lattice limit ApproxNEB barrier of 1015 meV for MnPO$_4$F is prohibitively high. Although MnPO$_4$F is an attractive cathode candidate from an energy density perspective, its poor solid-state Mg$^{2+}$ mobility would result in poor rate capability.

**MnO$_2$ (mp-25275)** is the spinel polymorph or $\lambda$ phase of MnO$_2$. Given that spinel LiMn$_2$O$_4$ is a well-studied, commercialized lithium-ion cathode,[46] there has been significant interest in studying spinels as Mg cathodes.[14,28,47] While promising from an energy density standpoint with a voltage of 2.9 V vs. Mg/Mg$^{2+}$ and theoretical capacity of 144 mAh/g, it has been challenging to experimentally realize a high capacity with repeated cycling because of sluggish Mg$^{2+}$ solid-state mobility.[34] High Mg$^{2+}$ migration barriers of ~800 meV for the dilute lattice limit (charged/deintercalated state) calculated with NEB have been reported for spinel MgMn$_2$O$_4$[14,28] which is consistent with the high (> 650 meV) migration barrier of 711 meV predicted by ApproxNEB in this screening method. This $\lambda$-MnO$_2$ spinel example demonstrates the value of the fourth screening tier where ApproxNEB is used to estimate migration barriers to identify materials where Mg$^{2+}$ solid-state mobility will be a challenge.

## Discussion

The migration pathways and relaxed ApproxNEB image structures were examined in more detail for the 4 highlighted Mg cathode candidates: NASICON V$_2$(PO$_4$)$_3$ (mp-26962), birnessite NaMn$_4$O$_8$ (mp-1016155), tavorite MnPO$_4$F (mp-25426), and spinel MnO$_2$ (mp-25275). In addition to mapping the energy difference along the pathway, the volume associated with the mobile Mg$^{2+}$ calculated using the Voronoi algorithm through pymatgen is included.[22,48–50] The graph area is colored to reflect the coordination of the Mg$^{2+}$ at various positions along the path based on the relaxed ApproxNEB images, where the coordination number was analyzed using the CrystalNN algorithm in pymatgen.[22,48] Images of the mobile Mg$^{2+}$ in the host crystal structure made using the VESTA software at significant points along the migration path are marked by letters and displayed with these plots in Figures 4-7.

A common feature in the pathways examined is the occurrence of high-energy (bottleneck) ion positions where the energy difference is highest (e.g. Figure 4 position B for sodiated birnessite NaMn$_4$O$_8$, Figure 5 position B for spinel MnO$_2$, Figure 6 position B & D for tavorite MnPO$_4$F, Figure 7 position B & D for NASICON V$_2$(PO4)$_3$) as the mobile Mg$^{2+}$ passes through a plane of anions. In the case where the Mg$^{2+}$ is moving from a tetrahedral to an octahedral site where the tetrahedra and octahedra are face sharing, this highest-energy point corresponds to the Mg$^{2+}$ squeezing through the triangle of anions composing the shared face. This finding is in line with previous work on spinels where the area of the anion triangle was expanded by substituting larger and more polarizable anion atoms in order to lower the energetic penalty for migration.[51] One counterintuitive finding, however, is that these bottleneck migration ion events do not correspond with the lowest volume sites along the migration path. In the identified pathways of these four materials, these lowest volume sites occur when Mg$^{2+}$ is in a 6-fold site of favored coordination, indicating a site with particularly favorable Mg-anion coordination and correspondingly tight bond lengths. This suggests that volume alone may not be a meaningful descriptor and more significant conclusions can be made by comparing volume across sites with similar local bonding environments.

The identified pathways for sodiated birnessite NaMn$_4$O$_8$ and spinel MnO$_2$ are both composed of face-sharing tetrahedra (tet) and octahedra (oct). While these two materials exhibit similar tet-oct-tet coordination changes, sodiated birnessite NaMn$_4$O$_8$ exhibits a much lower Mg$^{2+}$ ApproxNEB migration barrier (200meV) compared to spinel MnO$_2$ (711meV). One possible explanation is that the sodium in birnessite expands the interlayer spacing, reduces the electrostatic interaction between Mg$^{2+}$ and the oxygen layers, and increases the available volume along the pathway compared to spinel MnO$_2$. Therefore, the energy penalty for Mg$^{2+}$ migration is lowered because the Coulombic interactions between the mobile ion and host structure are weaker.

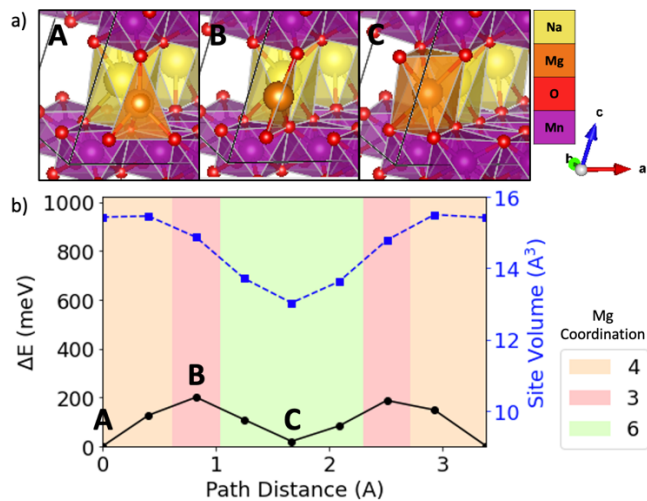

Figure 6. The evolving environment and associated Mg$^{2+}$ migration energy as a function of pathway coordinate in sodiated birnessite NaMn$_4$O$_8$ where b) depicts a graph of the energy barrier (black line with circles), Mg site volume (blue line with squares), and Mg$^{2+}$ coordination (colored graph area) based on ApproxNEB and a) shows images of the Mg$^{2+}$ at various positions along the migration pathway (labeled by A, B, and C).

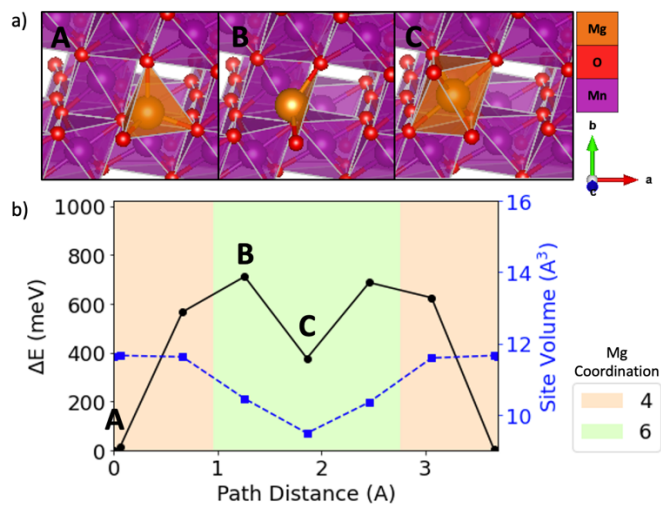

Figure 7. The evolving environment and associated Mg$^{2+}$ migration energy as a function of pathway coordinate in spinel MnO$_2$ where b) depicts a graph of the energy barrier (black line with circles), Mg site volume (blue line with squares), and Mg$^{2+}$ coordination (colored graph area) based on ApproxNEB and a) shows images of the Mg$^{2+}$ at various positions along the migration pathway (labeled by A, B, and C).

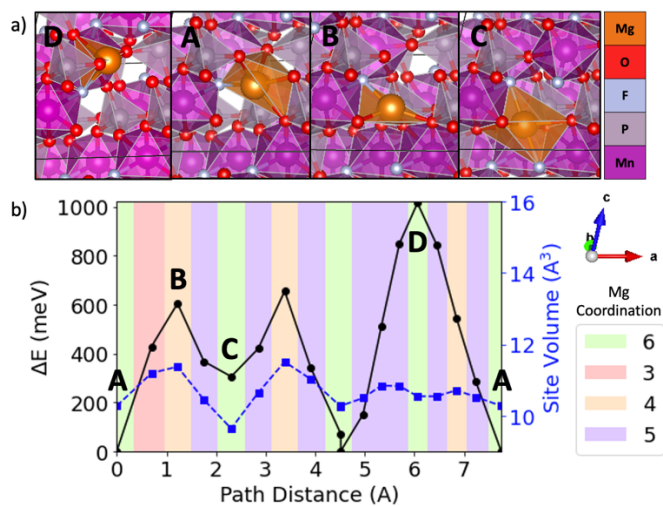

Figure 5. The evolving environment and associated Mg$^{2+}$ migration energy as a function of pathway coordinate in tavorite MnPO$_4$F where b) depicts a graph of the energy barrier (black line with circles), Mg site volume (blue line with squares), and Mg$^{2+}$ coordination (colored graph area) based on ApproxNEB and a) shows images of the Mg$^{2+}$ at various positions along the migration pathway (labeled by A, B, C, and D).

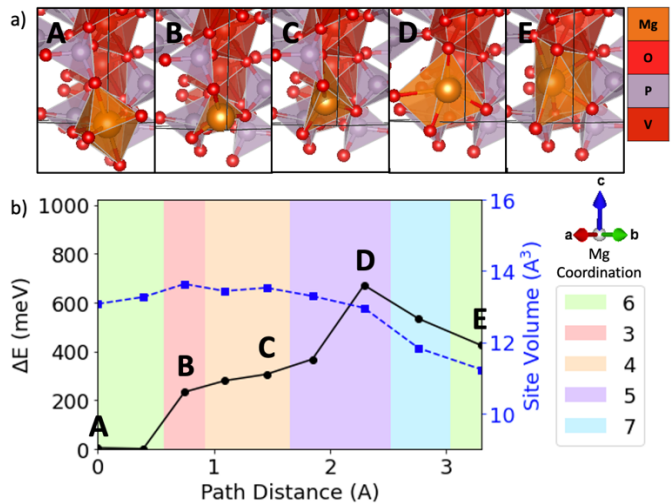

Figure 4. The evolving environment and associated Mg$^{2+}$ migration energy as a function of pathway coordinate in NASICON V$_2$(PO$_4$)$_3$ where b) depicts a graph of the energy barrier (black line with circles), Mg site volume (blue line with squares), and Mg$^{2+}$ coordination (colored plot area) based on ApproxNEB and a) shows the images of the Mg$^{2+}$ at various positions along the migration pathway (labeled by A, B, C, D, and E).

Examining the high energy positions in tavorite MnPO$_4$F illustrates the influence of anion composition when considering Mg$^{2+}$ passing through planes of anions. In Figure 6, the higher- energy bottleneck position D corresponds to the mobile Mg$^{2+}$ passing through a plane of 4 oxygen atoms. The Mg$^{2+}$ moves through a plane comprising of 2 oxygens and one fluorine at the lower energy bottleneck position B. At D, the Mg$^{2+}$ ion passes through the oxygen plane off center while at position B, the Mg$^{2+}$ ion is in the center of the anion triangle. Therefore, it is likely this position discrepancy contributes to the higher energy of point D, but it is also possible that substituting fluorine for one oxygen in the anion plane reduces the energetic penalty.

Of the 6 identified bottleneck positions where the mobile Mg$^{2+}$ ion passes through an anion plane in these 4 materials, only one case, point B of Figure 7 for NASICON V$_2$(PO$_4$)$_3$, does not correspond to a local energy maximum. We hypothesize that the relatively larger available volume for the mobile Mg$^{2+}$ in NASICON V$_2$(PO$_4$)$_3$ is responsible for the improved migration energetics, even when moving through anion planes. Both NASICON V$_2$(PO$_4$)$_3$ and sodiated birnessite $NaMn_4O_8$ exhibit large volume Mg sites (> 13 $Å^3$) as compared to spinel $MnO_2$ and tavorite $MnPO_4F$ (< 12 $Å^3$). The Mg sites exhibit larger volumes (> 13 $Å^3$) and relatively low energy differences (< 350 meV) when considering the NASICON $V_2(PO_4)_3$ oct-tet transition (Figure 7 A→B→C) and the sodiated birnessite $NaMn_4O_8$ tet-oct transition (Figure 4 A→B→C). Thus, the tetrahedra and octahedra volumes are large enough to avoid a substantial energy penalty when the $Mg^{2+}$ squeezes through the shared anion face. This suggests that perhaps there is a minimum anion triangle area where a costly energy penalty for $Mg^{2+}$ migration can be avoided.

## Conclusions

A new comprehensive materials screening framework designed to identify promising multivalent cathodes among materials that do not contain the active intercalating ion has been implemented and applied to Mg cathode discovery. The screening consists of four tiers which each focus on a different set of properties relevant for multivalent cathodes. In the first tier, all materials in the Materials Project were filtered by relative stability and composition. Then, candidate materials were filtered to include at least one reducible cation with a high enough oxidation state to permit intercalation. In the third tier, possible multivalent ion insertion sites were identified and candidates were prioritized by voltage, charge stability, and discharge stability. In the fourth and final screening tier, multivalent ion solid-state mobility is evaluated using the ApproxNEB algorithm.

From the reported work on Mg cathodes, four materials: NASICON $V_2(PO_4)_3$ (mp-26962), birnessite $NaMn_4O_8$ (mp-1016155), tavorite $MnPO_4F$ (mp-25426), and spinel $MnO_2$ (mp-25275) were identified as possible promising Mg cathodes and discussed in more detail. In these examples, different $Mg^{2+}$ migration barriers were observed despite having similar changes in coordination along the pathway. Local energy maxima were found to correlate with site topology (such as passing through an anion plane) rather than the lowest site volume along the path. However, when comparing only the anion plane sites, materials which exhibit a larger area between anions (such as NASICON $V_2(PO_4)_3$ and birnessite $NaMn_4O_8$) were found to reduce the energy penalty for $Mg^{2+}$ migration. Therefore, the available free volume for a given type of local bonding environment of a mobile ion site in the host structure was proposed as another influential factor for improving solid-state mobility.

## Author Contributions

## Conflicts of Interest

## Acknowledgements

This work was supported by the Volkswagen group. Materials data was provided by the Materials Project, which is funded by the U.S. Department of Energy, Office of Science, Office of Basic Energy Sciences, Materials Sciences and Engineering Division, under Contract No. DE-AC02-05-CH11231: Materials Project Program KC23MP. This work used computational resources provided by the National Energy Research Scientific Computing Center (NERSC), a U.S. Department of Energy Office of Science User Facility operated under Contract No. DE-AC02-05CH11231. The authors would also like to thank Gerbrand Ceder for helpful discussions that supported this work.

## Notes and References

1. Tian Y, Zeng G, Rutt A, et al. Promises and Challenges of Next-Generation "Beyond Li-ion" Batteries for Electric Vehicles and Grid Decarbonization. *Chem Rev*. 2021;121(3):1623-1669. doi:10.1021/acs.chemrev.0c00767
2. Liang Y, Dong H, Aurbach D, Yao Y. Current status and future directions of multivalent metal-ion batteries. *Nat Energy*. 2020;5(9):646-656. doi:10.1038/s41560-020-0655-0
3. Ponrouch A, Bitenc J, Dominko R, Lindahl N, Johansson P, Palacin MR. Multivalent rechargeable batteries. *Energy Storage Mater*. 2019;20:253-262.
4. Emly A, Van der Ven A. Mg Intercalation in Layered and Spinel Host Crystal Structures for Mg Batteries. *Inorg Chem*. 2015;54(9):4394-4402. doi:10.1021/acs.inorgchem.5b00188
5. Sun X, Bonnick P, Duffort V, et al. A high capacity thiospinel cathode for Mg batteries. *Energy Environ Sci*. 2016;9(7):2273-2277. doi:10.1039/C6EE00724D
6. Sun X, Bonnick P, Nazar LF. Layered TiS2 Positive Electrode for Mg Batteries. *ACS Energy Lett*. 2016;1(1):297-301. doi:10.1021/acsenergylett.6b00145
7. Aurbach D, Suresh GS, Levi E, et al. Progress in Rechargeable Magnesium Battery Technology. *Adv Mater*. 2007;19(23):4260-4267. doi:10.1002/adma.200701495
8. Ponrouch A, Frontera C, Bardé F, Palacín MR. Towards a calcium-based rechargeable battery. *Nat Mater*. 2016;15(2):169-172. doi:10.1038/nmat4462
9. Wang D, Gao X, Chen Y, Jin L, Kuss C, Bruce PG. Plating and stripping calcium in an organic electrolyte. *Nat Mater*. 2018;17(1):16-20. doi:10.1038/nmat5036
10. Li Z, Fuhr O, Fichtner M, Zhao-Karger Z. Towards stable and efficient electrolytes for room-temperature rechargeable calcium batteries. *Energy Environ Sci*. 2019;12(12):3496-3501. doi:10.1039/C9EE01699F
11. Xu Z-L, Park J, Wang J, et al. A new high-voltage calcium intercalation host for ultra-stable and high-power calcium rechargeable batteries. *Nat Commun*. 2021;12(1):3369. doi:10.1038/s41467-021-23703-x
12. Jeon B, Heo JW, Hyoung J, Kwak HH, Lee DM, Hong ST. Reversible calcium-ion insertion in NaSICON-type NaV2(PO4)3. *Chem Mater*. 2020;32(20):8772-8780. doi:10.1021/ACS.CHEMMATER.0C01112/SUPPL_FILE/CM0C01112_SI_001.PDF
13. Kim S, Yin L, Lee MH, et al. High-Voltage Phosphate Cathodes for Rechargeable Ca-Ion Batteries. *ACS Energy Lett*. 2020;5(10):3203-3211. doi:10.1021/acsenergylett.0c01663
14. Rong Z, Malik R, Canepa P, et al. Materials Design Rules for Multivalent Ion Mobility in Intercalation Structures. *Chem Mater*. 2015;27(17):6016-6021. doi:10.1021/acs.chemmater.5b02342
15. Aurbach D, Lu Z, Schechter A, et al. Prototype systems for rechargeable magnesium batteries. *Nature*. 2000;407(6805):724-727. doi:10.1038/35037553
16. Bölle FT, Mathiesen NR, Nielsen AJ, Vegge T, Garcia-Lastra JM, Castelli IE. Autonomous Discovery of Materials for Intercalation Electrodes. *Batter Supercaps*. 2020;3(6):488-498. doi:10.1002/batt.201900152
17. Gunter D, Cholia S, Jain A, et al. Community Accessible Datastore of High-Throughput Calculations: Experiences from the Materials Project. In: *2012 SC Companion: High Performance Computing, Networking Storage and Analysis*. IEEE; 2012:1244-1251. doi:10.1109/SC.Companion.2012.150
18. Ong SP, Wang L, Kang B, Ceder G. Li–Fe–P–O 2 Phase Diagram from First Principles Calculations. *Chem Mater*. 2008;20(5):1798-1807. doi:10.1021/cm702327g
19. Hautier G, Ong SP, Jain A, Moore CJ, Ceder G. Accuracy of density functional theory in predicting formation energies of ternary oxides from binary oxides and its implication on phase stability. *Phys Rev B*. 2012;85(15):155208. doi:10.1103/PhysRevB.85.155208
20. Sun W, Dacek ST, Ong SP, et al. The thermodynamic scale of inorganic crystalline metastability. *Sci Adv*. 2016;2(11). doi:10.1126/sciadv.1600225
21. Aykol M, Dwaraknath SS, Sun W, Persson KA. Thermodynamic limit for synthesis of metastable inorganic materials. *Sci Adv*. 2018;4(4). doi:10.1126/sciadv.aaq0148
22. Ong SP, Richards WD, Jain A, et al. Python Materials Genomics (pymatgen): A robust, open-source python library for materials analysis. *Comput Mater Sci*. 2013;68:314-319. doi:10.1016/j.commatsci.2012.10.028
23. O'Keefe M, Brese NE. Atom sizes and bond lengths in molecules and crystals. *J Am Chem Soc*. 1991;113(9):3226-3229. doi:10.1021/ja00009a002
24. Shen J, Horton M, Persson KA. A charge-density-based general cation insertion algorithm for generating new Li-ion cathode materials. *npj Comput Mater*. 2020;6(1):161. doi:10.1038/s41524-020-00422-3
25. Shen J-X, Li HH, Rutt AC, Horton MK, Persson KA. Rapid discovery of cathodes, ionic conductors and solid-state electrolytes through topological migration analysis. February 2022. doi:https://doi.org/10.48550/arXiv.2202.0022

26. Rong Z, Kitchaev D, Canepa P, Huang W, Ceder G. An efficient algorithm for finding the minimum energy path for cation migration in ionic materials. *J Chem Phys*. 2016;145(7):074112. doi:10.1063/1.4960790
27. Mathew K, Montoya JH, Faghaninia A, et al. Atomate: A high-level interface to generate, execute, and analyze computational materials science workflows. *Comput Mater Sci*. 2017;139:140-152. doi:10.1016/j.commatsci.2017.07.030
28. Liu M, Rong Z, Malik R, et al. Spinel compounds as multivalent battery cathodes: A systematic evaluation based on ab initio calculations. *Energy Environ Sci*. 2015;8(3):964-974. doi:10.1039/c4ee03389b
29. Dathar GKP, Sheppard D, Stevenson KJ, Henkelman G. Calculations of Li-Ion Diffusion in Olivine Phosphates. *Chem Mater*. 2011;23(17):4032-4037. doi:10.1021/cm201604g
30. Ong SP, Chevrier VL, Hautier G, et al. Voltage, stability and diffusion barrier differences between sodium-ion and lithium-ion intercalation materials. *Energy Environ Sci*. 2011;4(9):3680. doi:10.1039/c1ee01782a
31. Morgan D, Van der Ven A, Ceder G. Li Conductivity in LixMPO4 (M = Mn, Fe, Co, Ni) Olivine Materials. *Electrochem Solid-State Lett*. 2004;7(2):A30. doi:10.1149/1.1633511/XML
32. Wang A, Kingsbury R, McDermott M, et al. A framework for quantifying uncertainty in DFT energy corrections. *Sci Rep*. 2021;11(1):15496. doi:10.1038/s41598-021-94550-5
33. Huang Z-D, Masese T, Orikasa Y, Mori T, Yamamoto K. Vanadium phosphate as a promising high-voltage magnesium ion (de)-intercalation cathode host. *RSC Adv*. 2015;5(12):8598-8603. doi:10.1039/C4RA14416C
34. Canepa P, Sai Gautam G, Hannah DC, et al. Odyssey of Multivalent Cathode Materials: Open Questions and Future Challenges. *Chem Rev*. 2017;117(5):4287-4341. doi:10.1021/acs.chemrev.6b00614
35. Sun X, Duffort V, Mehdi BL, Browning ND, Nazar LF. Investigation of the Mechanism of Mg Insertion in Birnessite in Nonaqueous and Aqueous Rechargeable Mg-Ion Batteries. *Chem Mater*. 2016;28(2):534-542. doi:10.1021/acs.chemmater.5b03983
36. Nam KW, Kim S, Lee S, et al. The High Performance of Crystal Water Containing Manganese Birnessite Cathodes for Magnesium Batteries. *Nano Lett*. 2015;15(6):4071-4079. doi:10.1021/acs.nanolett.5b01109
37. Zhao Y, Han C, Yang J, et al. Stable Alkali Metal Ion Intercalation Compounds as Optimized Metal Oxide Nanowire Cathodes for Lithium Batteries. *Nano Lett*. 2015;15(3):2180-2185. doi:10.1021/acs.nanolett.5b00284
38. Liu H, Wang Y, Li L, Wang K, Hosono E, Zhou H. Facile synthesis of NaV6O15 nanorods and its electrochemical behavior as cathode material in rechargeable lithium batteries. *J Mater Chem*. 2009;19(42):7885. doi:10.1039/b912906e
39. Clites M, Pomerantseva E. Stabilization of battery electrodes through chemical pre-intercalation of layered materials. In: Kobayashi NP, Talin AA, Islam MS, Davydov A V., eds. *Low-Dimensional Materials and Devices 2016*. Vol 9924. SPIE; 2016:992405. doi:10.1117/12.2238655
40. Wang Q, Zhang M, Zhou C, Chen Y. Concerted Ion-Exchange Mechanism for Sodium Diffusion and Its Promotion in Na 3 V 2 (PO 4 ) 3 Framework. *J Phys Chem C*. 2018;122(29):16649-16654. doi:10.1021/acs.jpcc.8b06120
41. GOVER R, BURNS P, BRYAN A, SAIDI M, SWOYER J, BARKER J. LiVPO4F: A new active material for safe lithium-ion batteries. *Solid State Ionics*. 2006;177(26-32):2635-2638. doi:10.1016/j.ssi.2006.04.049
42. Chen D, Shao G-Q, Li B, et al. Synthesis, crystal structure and electrochemical properties of LiFePO4F cathode material for Li-ion batteries. *Electrochim Acta*. 2014;147:663-668. doi:10.1016/j.electacta.2014.09.131
43. Recham N, Chotard J-N, Dupont L, et al. A 3.6 V lithium-based fluorosulphate insertion positive electrode for lithium-ion batteries. *Nat Mater*. 2010;9(1):68-74. doi:10.1038/nmat2590
44. Wu J, Gao G, Wu G, et al. Tavorite-FeSO 4 F as a potential cathode material for Mg ion batteries: a first principles calculation. *Phys Chem Chem Phys*. 2014;16(42):22974-22978. doi:10.1039/C4CP03176H
45. Wu J, Gao G, Wu G, et al. MgVPO 4 F as a one-dimensional Mg-ion conductor for Mg ion battery positive electrode: a first principles calculation. *RSC Adv*. 2014;4(29):15014-15017. doi:10.1039/C4RA00199K
46. Dou S. Review and prospects of Mn-based spinel compounds as cathode materials for lithium-ion batteries. *Ionics (Kiel)*. 2015;21(11):3001-3030. doi:10.1007/s11581-015-1545-5
47. Kim C, Phillips PJ, Key B, et al. Direct Observation of Reversible Magnesium Ion Intercalation into a Spinel Oxide Host. *Adv Mater*. 2015;27(22):3377-3384. doi:10.1002/adma.201500083
48. Pan H, Ganose AM, Horton M, et al. Benchmarking Coordination Number Prediction Algorithms on Inorganic Crystal Structures. *Inorg Chem*. 2021;60(3):1590-1603. doi:10.1021/acs.inorgchem.0c02996
49. O'Keeffe M. A proposed rigorous definition of coordination number. *Acta Crystallogr Sect A Cryst Physics, Diffraction, Theor Gen Crystallogr*. 1979;35(5):772-775. doi:10.1107/S0567739479001765
50. Barber CB, Dobkin DP, Huhdanpaa H. The quickhull algorithm for convex hulls. *ACM Trans Math Softw*. 1996;22(4):469-483. doi:10.1145/235815.235821
51. Canepa P, Bo S-H, Sai Gautam G, et al. High magnesium mobility in ternary spinel chalcogenides. *Nat Commun*. 2017;8(1):1759. doi:10.1038/s41467-017-01772-1